\documentclass[aps,prx,twocolumn,longbibliography,superscriptaddress,floatfix,nofootinbib]{revtex4-2}
\usepackage{epsfig,amsmath,amssymb,color,comment,physics}
\usepackage[makeroom]{cancel}
\usepackage[caption=false]{subfig}
\usepackage{mathrsfs}
\usepackage[countmax]{subfloat}
\usepackage[normalem]{ulem}
\usepackage[english]{babel}
\usepackage{dsfont}
\usepackage{float}
\usepackage[bookmarks=true,colorlinks,linkcolor=black,urlcolor=NavyBlue,citecolor=RoyalBlue]{hyperref}
\usepackage{mathtools}
\usepackage{siunitx}
\usepackage{upgreek}
\usepackage{comment}
\usepackage{xcolor}

\hypersetup{
    unicode=false,     
    pdftoolbar=false,  
    pdfmenubar=true,   
    pdffitwindow=false, 
    pdfstartview={FitH},
    pdftitle={},    
    pdfauthor={Authors},     
    pdfsubject={},   
    pdfcreator={},   
    pdfproducer={}, 
    pdfkeywords={quantum many-body scars} {superconducting processor} {quantum state tomography}, 
    pdfnewwindow=true,
    colorlinks=true,
    linkcolor=black,
    citecolor=blue, 
    filecolor=magenta,
    urlcolor=blue
}

\newcommand{\Er}{\textrm{E}_{\textrm{r}}}

\newcommand{\mus}{\upmu\textrm{s}}

\begin{document}

\title{Interaction-enabled metal-insulator phase transition in a driven quantum gas 
}

\author{ Camilo Cantillano } \thanks{These authors contributed equally to this work.}
\affiliation{Institut f{\"u}r Experimentalphysik und Zentrum f{\"u}r Quantenphysik, Universit{\"a}t Innsbruck, Technikerstra{\ss}e 25, Innsbruck, 6020, Austria} 

\author{ Karthick Ramanathan } \thanks{These authors contributed equally to this work.}
\affiliation{Institut f{\"u}r Experimentalphysik und Zentrum f{\"u}r Quantenphysik, Universit{\"a}t Innsbruck, Technikerstra{\ss}e 25, Innsbruck, 6020, Austria} 

\author{ Zekai Chen }
\thanks{These authors contributed equally to this work.}
\affiliation{Department of Physics and Astronomy, University of Rochester, Rochester, New York, 14627, USA}
\affiliation{Institut f{\"u}r Experimentalphysik und Zentrum f{\"u}r Quantenphysik, Universit{\"a}t Innsbruck, Technikerstra{\ss}e 25, Innsbruck, 6020, Austria} 

\author{ Ang Yang }
\affiliation{School of Physics and Zhejiang Key Laboratory of Micro-nano Quantum Chips and Quantum Control, Zhejiang University, Hangzhou 310027, China} 

\author{ Emilio  Aguilera-Valdes }
\affiliation{Institut f{\"u}r Experimentalphysik und Zentrum f{\"u}r Quantenphysik, Universit{\"a}t Innsbruck, Technikerstra{\ss}e 25, Innsbruck, 6020, Austria} 

\author{ Lei Ying}
\email{leiying@zju.edu.cn}
\affiliation{School of Physics and Zhejiang Key Laboratory of Micro-nano Quantum Chips and Quantum Control, Zhejiang University, Hangzhou 310027, China}

\author{ Manuele  Landini}
\affiliation{Institut f{\"u}r Experimentalphysik und Zentrum f{\"u}r Quantenphysik, Universit{\"a}t Innsbruck, Technikerstra{\ss}e 25, Innsbruck, 6020, Austria}

\author{ Hanns-Christoph  N{\"a}gerl}\email{christoph.naegerl@uibk.ac.at}
\affiliation{Institut f{\"u}r Experimentalphysik und Zentrum f{\"u}r Quantenphysik, Universit{\"a}t Innsbruck, Technikerstra{\ss}e 25, Innsbruck, 6020, Austria}

\author{ Yanliang  Guo }\email{yanliang.guo@uibk.ac.at}
\affiliation{Key Laboratory of Quantum State Construction and Manipulation (Ministry of Education), School of Physics, Renmin University of China, Beijing 100872, China}
\affiliation{Institut f{\"u}r Experimentalphysik und Zentrum f{\"u}r Quantenphysik, Universit{\"a}t Innsbruck, Technikerstra{\ss}e 25, Innsbruck, 6020, Austria}

\date{\today}

\begin{abstract}

Particle transport and energy flow are central to a wide range of phenomena in the natural sciences. While interactions generically promote ergodicity and diffusion~\cite{PhysRevA.43.2046,PhysRevE.50.888,rigol2008thermalization}, quantum interference can arrest transport, defying classical expectations~\cite{PhysRev.109.1492,RevModPhys.91.021001}. Here, we experimentally investigate their interplay in a periodically driven 3D quantum gas~\cite{Bukov04032015,RevModPhys.89.011004} with tunable interactions. Strikingly, we find a sharp dynamical boundary separating localization from diffusive energy absorption. By tuning the driving amplitude and interaction strength, we map the localization-delocalization phase diagram and characterize this boundary via finite-time scaling. On the insulating side, we observe many-body dynamical localization (MBDL)~\cite{guo_observation_2023} featuring arrested momentum-space transport. Transport becomes subdiffusive near the boundary and diffusive in the delocalized regime, yielding a metal-insulator transition that we interpret as localization in many-body Hilbert space. Our results exemplify an interaction-enabled dynamical phase transition in a closed Floquet many-body system, and clarify how coherence and interactions jointly govern the quantum-to-classical transition.

\end{abstract}
\maketitle
Understanding how interacting many-particle systems relax and transport energy under external perturbations is a central question in many fields of science. At a fundamental level, this question poses a challenge because it generally concerns non-equilibrium dynamics that is not fully captured by traditional statistical physics based on assumptions of equilibrium and weak interactions~\cite{RevModPhys.83.863}. In classical systems, interactions generically promote diffusion and thermalization, providing a robust route towards equilibrium. In contrast, quantum mechanics modifies this expectation in a fundamental way~\cite{PhysRevA.43.2046,PhysRevE.50.888,rigol2008thermalization}. Coherent interference can suppress transport, producing localization and the breakdown of ergodic behavior~\cite{PhysRev.109.1492}. Understanding when interactions restore diffusion and when interference prevails remains a central problem in modern physics.

Periodically driven systems provide a powerful setting to probe the competition between interactions and coherence. Continuous driving injects energy into the system~\cite{Haake2019-jy,Chirikov1979chaos}, promoting diffusive ergodic transport and leading to classical behavior~\cite{PhysRevX.4.041048,PhysRevE.90.012110,PONTE2015196,Bukov04032015}. A paradigmatic counterexample is dynamical localization (DL) in the quantum kicked rotor (QKR)~\cite{Casati1979,santhanam_quantum_2022,Sieberer2019-mz}, where quantum interference suppresses diffusion in momentum space~\cite{PhysRevLett.75.4598}. In the non-interacting limit, the system maps to a one-dimensional (1D) Anderson model~\cite{PhysRevLett.49.509}, and energy absorption is suppressed for all parameters. Richer behavior emerges for quasi-periodic driving, which effectively increases the dimensionality of the Anderson model and gives rise to a metal-insulator transition in non-interacting systems~\cite{PhysRevA.80.043626,madani_observation_2025}. Beyond this, interacting systems have been found to exhibit subdiffusive energy growth, suggesting that interactions can restore transport~\cite{cao2022interaction,PhysRevLett.100.094101,gligoric_interactions_2011}. Recently, MBDL has been observed in the quasi-integrable 1D regime~\cite{guo_observation_2023}, where ergodic mixing remains limited. Together, these results highlight the role of interference, effective dimensionality, and interactions in shaping transport in driven systems. Yet, a central question remains unresolved: in a driven isolated 3D many-body system, can interactions alone generate a sharp boundary between bounded and diffusive transport? 

\begin{figure*}
\centering
\renewcommand{\figurename}{Fig}
\includegraphics[width=0.9\linewidth]{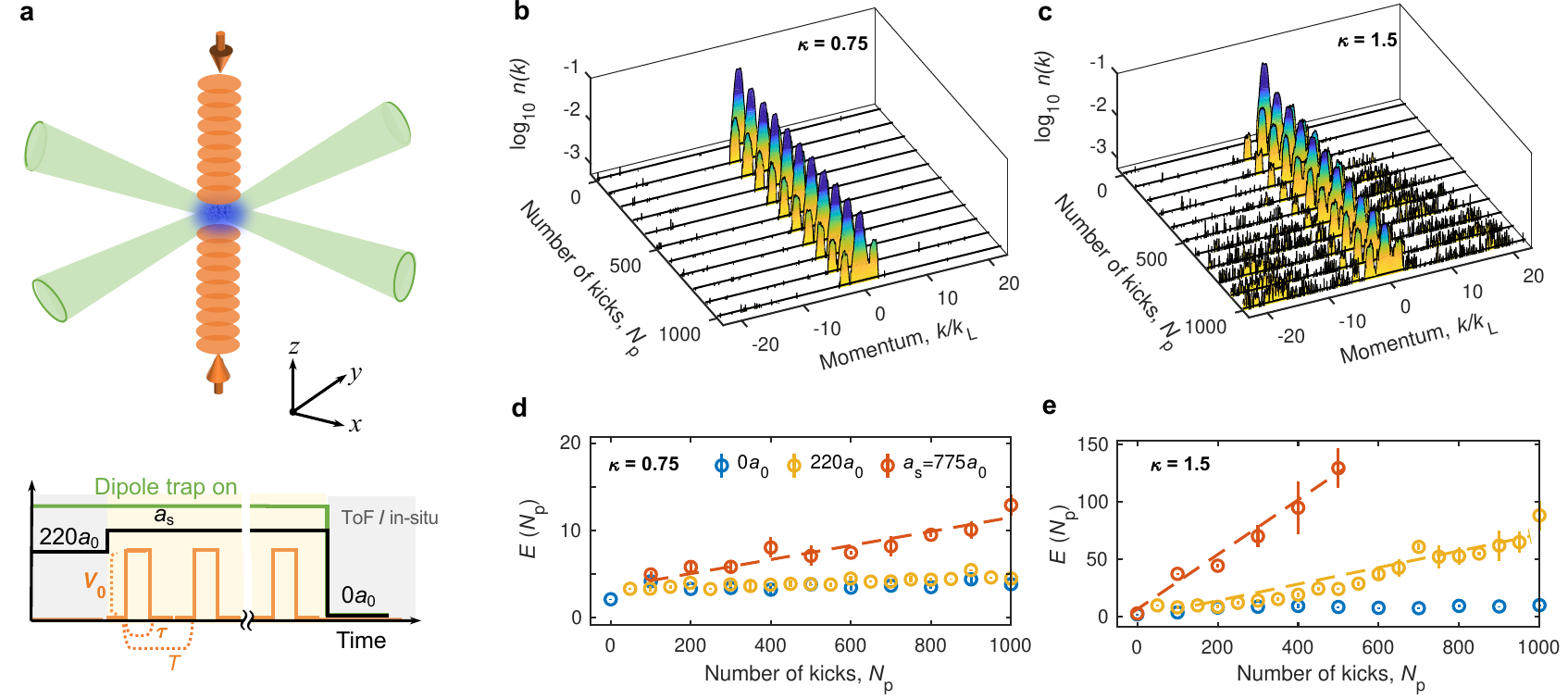}
\caption{{\bf Experimental setup and observation of localization and delocalization.} 
$\bf{a}$ Schematics of the experimental setup (top) and of the protocol (bottom): The BEC (blue) confined in crossed trapping beams (green) is subjected to $N_\mathrm{p}$ pulses of a vertically oriented optical lattice (orange) before undergoing ToF imaging or, alternatively, in-situ imaging. $\bf{b, c}$, Momentum distribution $n(k)$ plotted after applying a certain number of kicks $N_\mathrm{p}$ for (b) $\kappa \!=\! 0.75, a_\mathrm{s} \!=\! 220\,a_0$ and (c) $\kappa \!=\! 1.5, a_\mathrm{s} \!=\! 775\,a_0$. $\bf{d, e}$, Energy proxy $E$ as a function of $N_\mathrm{p}$ for 3 values of the scattering length $a_s\!=\!0 \, a_0$ (blue circles), $220 \, a_0$ (yellow circles), and $775 \, a_0$ (red circles) for (d) $\kappa \!=\! 0.75$ and (e) $\kappa \!=\! 1.5$. The dashed lines indicate linear fits. All experimental data are the average of three realizations.}
\label{fig1}
\end{figure*}

In this work, we report on a direct experimental investigation of this scenario by measuring the transport dynamics of a periodically driven, interacting 3D quantum gas. By varying both the kick strength and the interparticle interaction strength, we map the transport behavior of the system in momentum space, revealing a localization-delocalization phase diagram. We characterize the phase transition with finite-time scaling analysis, observing universal features typical of a metal-insulator transition. We switch dynamically between the two phases, reversibly traversing the transition point, which supports a second-order phase transition scenario. In addition, measurements of the in-situ spatial distribution reveal that delocalization in momentum space is accompanied by diffusion in real space, while localization in momentum space corresponds to the fact that the atoms remain trapped. We give an interpretation in terms of localization in many-body Hilbert space~\cite{PhysRevLett.78.2803}, where the wave function remains confined to a small region, despite interactions and driving.

For the experiment, we prepare a nearly pure Bose-Einstein condensate (BEC) of about $\mathcal{N}_0\!=\!10^4$ $^{133}$Cs atoms in a crossed dipole trap~\cite{Kraemer2004, Supp}, as illustrated in Fig.~\ref{fig1}a, levitated against gravity by a magnetic field gradient. With the magnetic offset field $B$ we tune the interatomic interaction by means of a Feshbach resonance ~\cite{FeshbachGrimm} for the $s$-wave scattering length $a_s(B)$. We set $a_s$ to values from 0(1.7) to 1037(1) $a_0$, where $a_0$ is the Bohr radius. The tunable QKR is implemented by periodically pulsing a standing wave that propagates in the vertical $z$-direction with lattice spacing $a\!=\!\pi/k_\text{L}\!=\!532.2 \,\textrm{nm}$ and lattice depth $V_0\!=\!30(1.5)$ to $84(1.5)$ $\Er$. Here, $k_\text{L}$ is the wave number and $\Er\!=\!\pi^2\hbar^2/(2ma^2)$ is the photon recoil energy. We choose square-pulses with a duration of $\tau\!=\!5.0$ $\mus$ and a period of $T\!=\!31.0$ $\mus$, as shown in Fig.~\ref{fig1}a. The (dimensionless) kick strength relevant to the QKR model is the pulse area in units of Planck's constant $\hbar$, i.e., $\kappa\!=\! V_0 \tau / 2\hbar$~\cite{PhysRevA.80.043626}. After applying a desired number of pulses $N_\mathrm{p}$, we turn off the interaction and the trap and either take an in-situ absorption image along the $x-y$ direction perpendicular to $z$, or perform a levitated time-of-flight (ToF) expansion with duration $t_{\textrm{\tiny ToF}}\!=\!12$ ms before taking the absorption image. In the latter case, integrating over the remaining transverse direction gives the momentum distribution $n(k)$ along the kick direction. We scale $n(k)$ by the initial atom number $\mathcal{N}_0$. Note that alternatively normalizing the evolved distributions by the instantaneous atom number would erase information about fast atoms potentially leaving the ToF observation window. In addition, to ascertain the interacting nature of our QKR system after it has evolved for $N_\mathrm{p}$ kicks, we carry out interaction-quench tests~\cite{Supp}.

\begin{figure*}
\centering
\renewcommand{\figurename}{Fig}
\includegraphics[width=0.72\linewidth]{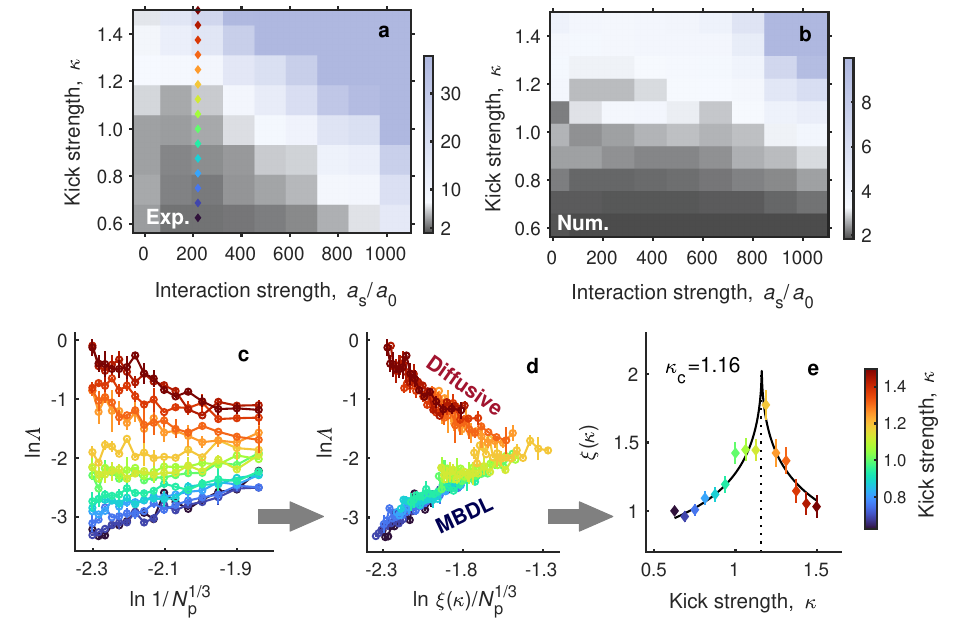}
\caption{{\bf Metal-insulator phase diagram and finite-time scaling analysis.}
{\bf a,b}, Phase diagrams constructed from the measured ({\bf a}) and the numerically determined ({\bf b}) energy proxy $E(N_\mathrm{p})$ after $N_\mathrm{p}\!=\!500$ kicks as a function of $\kappa$ and $a_s$.
{\bf c}, Scaling function $\ln \Lambda$ as a function of $\ln[1/N_\mathrm{p}^{1/3}]$ for different values of $\kappa$ as measured at $a_s\!=\!220\,a_0$.
{\bf d}, Data collapse in (c) onto a universal scaling function, with $\ln \Lambda$ plotted against the scaled quantity $\ln[\xi(\kappa)/N_k^{1/3}]$. {\bf e}, Extracted scaling parameter $\xi(\kappa)$ as a function of $\kappa$ with a power-law fit (solid black line). The critical value of $\kappa_c \!=\! 1.16$ is indicated by the dotted line.}
\label{fig2}
\end{figure*}

We first study the time evolution of $n(k)$ under periodic driving up to $N_\mathrm{p}\!=\!1000$ kicks at selected values of $a_s$ and $\kappa$. The data are shown in Fig.\ref{fig1}b and c. Prior to kicking, $n(k)$ is singly peaked at zero momentum with an rms-width of $0.50(2)k_\text{L}$. Upon the first kicks, it develops prominent side peaks at $\pm 2k_\mathrm{L}$. For moderate kick and interaction strengths, $\kappa\!=\!0.750(3)$ and $a_s \!=\! 220(1.2)\,a_0$, the distribution subsequently reaches a stationary state. Evidently, the system is localized even in the presence of interactions. The situation is very different at higher kick and interaction strengths for which we observe transport to higher momentum states. Now the system becomes delocalized. To characterize the dynamical evolution, we compute the quantity $E(N_\mathrm{p}) \!=\! \abs{\int_{-0.5k_\text{L}}^{0.5k_\text{L}} n(k) dk}^{-2}$, which can be viewed as an energy proxy~\cite{PhysRevA.80.043626}. Its value increases as atoms leave the zero-momentum peak and is therefore sensitive to momentum-space transport even when the finite imaging window prevents a direct extraction of the full kinetic energy. Its time evolution is plotted in Fig.\ref{fig1}d and e. For zero interactions, $E(N_\mathrm{p})$ saturates after a short evolution to a relatively low value. For strong interactions ($a_s \!=\! 775\,a_0$) it shows linear (diffusive) growth. For intermediate interactions ($a_s \!=\! 220\,a_0$) and moderate kick strength ($\kappa\!=\!0.75$) its arrested evolution reflects the localization behavior that is seen from the evolution of $n(k)$. Evidently, there are two distinct transport regimes depending on the interaction strength at a given kick strength: saturation signifying DL or MBDL at zero or moderate interactions, respectively, and monotonic growth corresponding to delocalization for sufficiently strong interactions. Our results indicate the presence of an MBDL-to-delocalization phase transition enabled by interactions. The position of the phase boundary depends on $\kappa$ and $a_s$.

We now aim to investigate the robustness of the MBDL phase and to characterize the phase transition to the delocalized regime. We measure $E(N_\mathrm{p})$ at $N_\mathrm{p}\!=\!500$ kicks for various values of $\kappa\!\in\![0.62, 1.75]$ and of $a_s\!\in\![0, 1037]\,a_0$. This allows us to construct a phase diagram as shown in Fig.~\ref{fig2}a. Localization is found in the left-hand corner for weak and intermediate interaction strengths and moderate kick strengths. Interestingly, MBDL occurs for non-zero interactions over a comparatively large range of kick strengths. Stronger interactions lead to lower critical kick strengths for the transition to the delocalized regime. As shown in Fig.~\ref{fig2}b, large-scale-simulation results based on the Gross-Pitaevskii equation (GPE)~\cite{Supp} are in reasonable qualitative agreement with the experimental data. Furthermore, to identify the transition line, we employ a one-parameter finite-time scaling analysis~\cite{PhysRevLett.42.673, PhysRevA.80.043626, olsen_interaction_2025}. The  dynamics of the system can be described by the scaling theory of the Anderson transition~\cite{wegner_electrons_1976} under the hypothesis that the energy evolution follows a universal function of a unique scaling variable. The critical boundary of the phase diagram is governed by sub-diffusive energy growth $E \propto N_\mathrm{p}^\beta$, where $\beta\!\in\!(0,1)$ is the exponent of the time evolution at criticality. To locate the transition point, we scan the value of $\kappa$ between 0.62-1.5 at fixed $a_s\!=\!220\,a_0$, as indicated by the colored points in Fig.\ref{fig2}a. The scaling function is defined as
\begin{equation}
    \Lambda (\kappa,N_\mathrm{p}) = E(N_\mathrm{p})\cdot N_\mathrm{p}^{-\beta} = f\left[\xi(\kappa)/N_\mathrm{p}^{1/d}\right],
    \label{eq:scaling_funct}
\end{equation}
where $f$ is the universal function, $\xi(\kappa)$ is the scaling parameter proportional to the localization length, and $d$ is an effective dimension~\cite{Supp}, which we expect to depend on how interactions modify the underlying structure of the Hilbert space~\cite{yang2025originemergentfeaturesmanybody,olsen_interaction_2025}. We consider values of $N_\mathrm{p}$ up to 1000 to capture the long-term dynamics. As shown in Fig.\ref{fig2}c, the unscaled evolution of $\ln \Lambda$ reveals a clear bifurcation at large $N_\mathrm{p}$: sub-critical trajectories bend downward, reflecting energy saturation, while super-critical curves bend upward, signaling diffusion. Motivated by the fact that Anderson-type localization transitions require an effective dimension larger than $d\!\!=\!\!2$, we analyze the data using $d\!=\!3$~\cite{Supp}. Residual minimization collapses these distinct trajectories onto two universal branches that represent the MBDL phase and the delocalized phase, respectively, as can be seen in Fig.\ref{fig2}d. This collapse is achieved by determining the time shifts, $\ln \xi(\kappa)$, required to align the trajectories. A power-law fit to the scaling parameter $\xi(\kappa)$ yields a critical kick strength of $\kappa_\text{c}\!=\!1.162(13)$ \cite{Supp}. The scale-invariant behavior observed here and the diverging localization length indicate that the transition from MBDL to delocalization is a second-order quantum phase transition. 

\begin{figure}
\centering
\renewcommand{\figurename}{Fig}
\includegraphics[width=\linewidth]{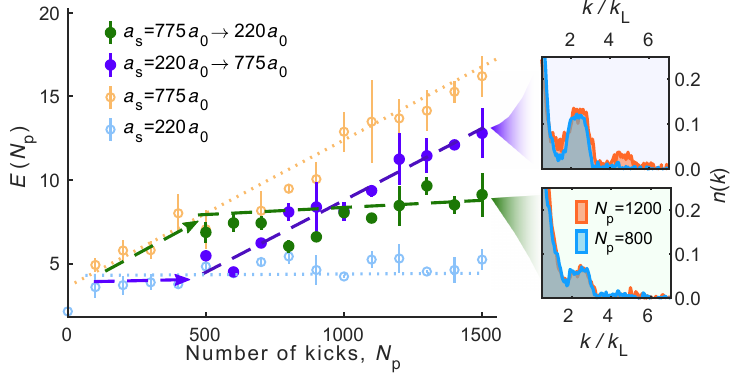}
\caption{{\bf Reversibility of the MBDL-to-delocalization phase transition.}
Evolution of $E$ versus $N_\mathrm{p}$ at $\kappa\!=\!0.75$ for $a_s\!=\!220\,a_0$ (blue circles), $775\,a_0$ (orange circles), and interaction ramp-up from $220\,a_0$ to $775\,a_0$ (purple dots) and ramp-down from $775\,a_0$ to $220\,a_0$ (green dots) at $N_\mathrm{p}\!=\!400$. The dashed and dotted lines are linear fits. The insets provide a comparison of $n(k)$ at $N_p\!=\!800$ (blue line) and $N_p\!=\!1200$ (red line) for the loss of MBDL (case A), top) and the regaining of MBDL (case B), bottom).}
\label{fig3}
\end{figure}

\begin{figure*}
\centering
\renewcommand{\figurename}{Fig}
\includegraphics[width=0.95\linewidth]{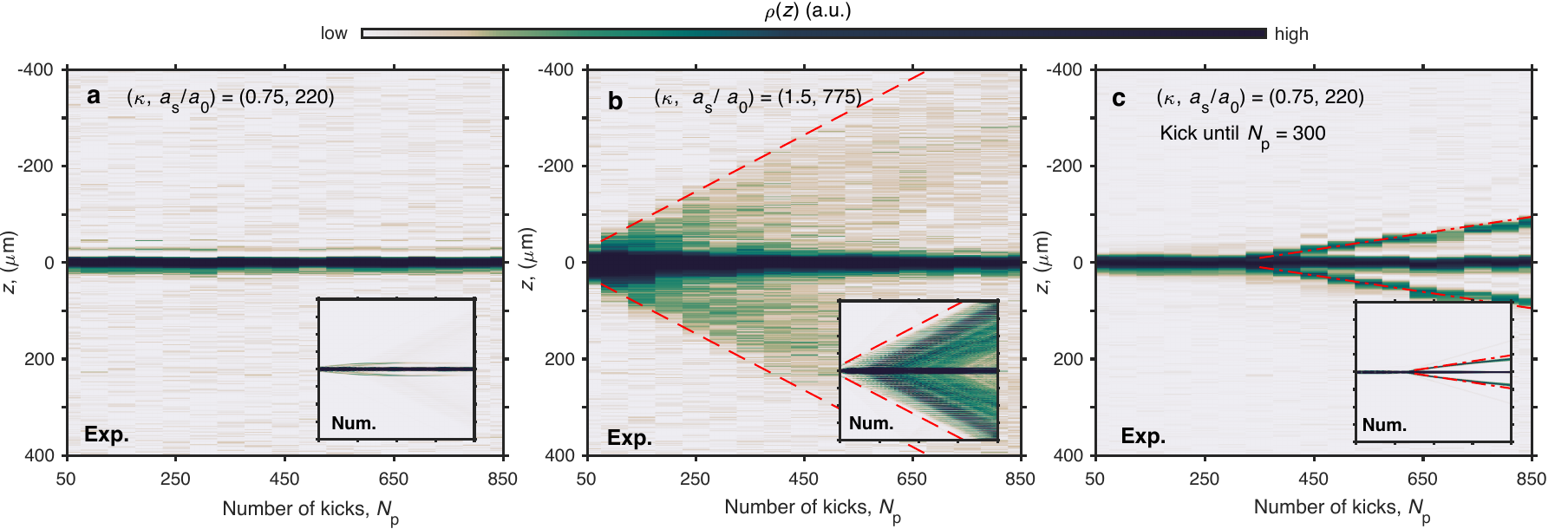}
\caption{
{\bf Real-space dynamics for many-body QKR.} Time evolution of the spatial density $\rho(z)$ for {\bf (a)} the MBDL case with ($\kappa, a_s/a_0$) = ($0.75$, $220$), for {\bf (b)} the delocalized case with ($\kappa, a_s/a_0$)=($1.5$,$775$), and for {\bf (c)} the MBDL case for which the kicking is stopped at $N_p\!=\!300$. Insets: Numerical simulation results under the same conditions as the experiment. The red dashed lines in {\bf (b)} and the dashed-dotted lines in {\bf (c)} indicate linear expansion for a speed corresponding to $\pm 6.5\hbar k_\text{L}$ and $\pm 2\hbar k_\text{L}$, respectively, as a guide to the eye.  
}
\label{fig4}
\end{figure*}

Next, we study how the transition is traversed as interactions are varied. For this, we implement interaction ramps during the kicking protocol. We initialize the system in either the MBDL or the delocalized phase and apply $N_\mathrm{p} \!=\! 400$ kicks. We then ramp the interaction strength to a new value over $3$ ms while continuing to kick. We compare the resulting dynamics against reference localized and delocalized cases, where the interactions are kept constantly low or high, respectively. Figure \ref{fig3} illustrates the evolution of the energy proxy $E$ for two exemplary cases. In case A, when ramping from low to high interactions, $E$ departs from its saturated value as the system transitions into the delocalized phase. The subsequent evolution of the momentum distribution $n(k)$ at $N_\mathrm{p} \!=\! 800$ and 1200 clearly reflects this breakdown of MBDL (upper inset). Conversely, in case B, when ramping from high to low interactions, the growth of $E$ abruptly halts and effectively saturates. Strikingly, the system regains MBDL independent of the energy accumulated during the initial evolution. The subsequent arrested evolution of $n(k)$, shown in the lower inset, confirms this restoration of localization. Evidently, the phase transition is reversible, providing direct evidence that it is governed by intrinsic many-body dynamics rather than irreversible heating.

We finally address the question of real-space dynamics. It is not a priori obvious how the spatial density $\rho(z)$ along the kick direction would evolve in view of the generation of non-zero momentum components not only in the delocalized, but also in the localized phase. Note that the trap has a depth of about $3.0\Er$, while the components at $\pm 2k_\mathrm{L}$ carry a kinetic energy of $4.0\Er$. The results of in-situ measurements are shown for two exemplary cases in Fig.~\ref{fig4}a and b. In the first case for moderate kick strength and moderate interaction strength, for which the momentum-space evolution (cf. Fig.~\ref{fig1}b) shows clear localization, we find that the spatial distribution also remains frozen. In contrast, for the second case (cf. Fig.~\ref{fig1}c), in the momentum-space delocalized regime, the real-space sample exhibits clear expansion. Note that in this case, the ToF measurement does not reflect the exact momentum distribution of the system. We further investigate the real-space dynamics by initializing the system in the MBDL regime with $\kappa\!=\!0.75$ and $a_s\!=\!220a_0$, and then halting the kicking sequence at $N_p\!=\!300$, see Fig.~\ref{fig4}c. After the drive is switched off, the density distribution evidently fragments, with distinct side peaks separating from the central component with momenta of about $\pm2\hbar k_L$. This reveals that the localized state contains finite-momentum components, but that under continued driving these components remain coherently bound in the Floquet dynamics rather than producing ballistic expansion. In all cases, the experimental observations align well with the results of the numerical simulations. These observations suggest that, in the dynamically localized regime, continued kicking in the presence of a trap constrains not only momentum-space transport but also supports real-space confinement.

In conclusion, we experimentally identify a sharp dynamical boundary that separates bounded from diffusive energy absorption in a periodically driven, interacting quantum gas. The MBDL phase~\cite{guo_observation_2023,PhysRevLett.124.155302} persists across an extended parameter range beyond quasi-integrable regimes, and the resulting localization-delocalization phase transition is captured by finite-time scaling analysis which reveals critical behavior with an effective dimension larger than one~\cite{PhysRevLett.42.673,PhysRevA.80.043626,cherroret_how_2014}.  The reversibility of the transition and the emergence of scale-invariant dynamics support an interpretation in terms of an interaction-enabled dynamical phase transition in an effectively isolated system,  in the absence of an external bath~\cite{ammann_quantum_1998, klappauf_observation_1998, steck_quantitative_2000, darcy_quantum_2001, schomerus_nonexponential_2007}. Interactions in a driven quantum gas do not merely wash out interference~\cite{cao2022interaction,shepelyansky_delocalization_1993,gligoric_interactions_2011}. Instead, they reorganize the accessible many-body configuration space~\cite{yang2025originemergentfeaturesmanybody,olsen_interaction_2025} and generate a tunable dynamical boundary between localized and diffusive transport. 
This picture connects naturally to localization in Fock space~\cite{PhysRevLett.78.2803,10.1063/1.459677,dupont_many-body_2026}, which underlies many-body localization (MBL) in disordered systems~\cite{BASKO20061126,annurev:/content/journals/10.1146/annurev-conmatphys-031214-014701,RevModPhys.91.021001}. In contrast to conventional MBL, our system does not rely on quenched disorder. Instead localization emerges dynamically from the interplay of interference and interactions in a driven setting. This makes the system a controllable platform for studying how classical diffusion emerges from coherent many-body dynamics~\cite{PhysRevX.4.041048,PhysRevE.90.012110,Bukov04032015}.

Understanding how the effective dimension depends on interaction strength, and whether it can be tuned continuously, remains an important open direction. This perspective connects to broader questions of dimensionality and universality. Exploring dimensional crossovers, for example, by tightening the transverse confinement toward a strictly 1D geometry, offers a direct route to testing how real-space dimensionality constrains many-body transport. The observation of real-space dynamics offers a potentially useful new perspective for understanding how Floquet interference shapes transport in the kicked rotor, particularly in the interacting many-body regime. Finally, the role of energy and initial conditions remains largely unexplored. Higher temperatures may enhance interaction-induced mixing and reshape the phase boundary, raising the possibility of energy-dependent transport regimes reminiscent of mobility edges~\cite{BASKO20061126,RevModPhys.91.021001}. More broadly, our results motivate the study of how classical transport may emerge intrinsically from interactions in an isolated quantum system.

\bigskip

\noindent{\bf Acknowledgments}\\
The Innsbruck team acknowledges funding by the Tyrolean Science Fund (TWF) under project number WF-F.50270/16-2025, by the European Research Council (ERC) under project number 101201611, and by an FFG infrastructure grant with project number FO999896041. Y.G. is supported by the Austrian Science Fund (FWF) with project number 10.55776/COE1, and Quantum Science and Technology-National Science and Technology Major Project of China (Grant No.2025ZD0300400). L.Y. and A.Y are supported by the National Natural Science Foundation of China (Grant No. 12375021), the Zhejiang Provincial Natural Science Foundation of China (Grant No. LD25A050002), and the National Key Research and Development Program of China (Grant No. 2022YFA1404203).  
\textbf{Author Contributions:} The work was conceived by Y.G., H.C.N. and M.L. Experiments were prepared and performed by C.C., K.R, Z.C. and E.A. Data were analyzed by C.C, K.R, E.A., A.Y, Y.G. and Z.C. Numerical simulations were performed by Z.C., A.Y. and L.Y. The manuscript was drafted mainly by Y.G., Z.C., M.L., K.R., C.C., L.Y. and H.C.N. All authors contributed to the discussion and finalization of the manuscript. 
{\bf Data Availability:} The data shown in the main text are available via Zenodo~\cite{Zenodo}.
{\bf Code Availability:} Codes supporting the findings of this study are available from the corresponding author on a reasonable request.


\cleardoublepage
\section{Supplemental materials}\label{sec:Supp}

\tableofcontents

\subsection{Experimental parameters}
Our experiment begins with $\mathcal{N}_0\!=\!1\!\times\!10^4$ atoms in a nearly pure BEC of $^{133}$Cs atoms prepared in their lowest hyperfine state $\ket{F,m_F}\!=\!\ket{3,3}$. The BEC is held in a crossed dipole trap with trap frequencies $(\omega_x,\omega_y,\omega_z)\!=\!2\pi\times(11.8(1),  15.5(2), 18.7(3))\,$Hz and levitated against gravity by a magnetic field gradient $\partial B/\partial z\! \!=\!31.07\,$G/cm~\cite{Kraemer2004}. We estimate the trap depth along the vertical $z$-direction to $189(8)$ nK, which is $2.97(13) \Er$. The BEC is initially prepared in the Thomas-Fermi regime at $a_s\approx 220\,$$a_\mathrm{0}$. The interactions are tuned to the desired value by means of a Feshbach resonance. The initial peak density is estimated to $9.7\times10^{12} \, \textrm{cm}^{-3}$, and the Thomas-Fermi radii are $ (10.9, 8.3, 6.9)\, \mu$m. The experimental phase diagrams (Fig.~\ref{fig2}a and Fig.~\ref{fig4}c) are determined by measuring the energy proxy at 8 equally spaced kick strengths $\kappa$ ranging from 0.62 to 1.5, for each of nine fixed scattering lengths: $a_s \!=\! 0$, $112$, $220$, $350$, $463$, $611$, $775$, $920$, and $1037\,a_0$.\\

\subsection{Numerical model and parameters}
In our numerical simulation, we consider a finite pulse length $\tau$ for the QKR instead of the ideal $\delta$-kicks. We therefore describe the system by the Hamiltonian~\cite{gligoric_interactions_2011, Delande2020meanfield,cao2022interaction}
\begin{eqnarray}\label{eq_H}
    \mathcal{H}(t)=\frac{\mathbf{P}^2}{2m}&+&\mathrm{V}(\mathbf{R})+g{\abs{\psi(\mathbf{R},t)}}^2 \nonumber \\
    &+&V_{\textrm{kick}}\sum_{n=1}^{N_\mathrm{p}}f(t,\tau,N_pT),
\end{eqnarray}
where $f(t,\tau,nT) \!\!=\!\!\vartheta(t-nT)\!-\!\vartheta(t-\tau-nT)$, $\vartheta(t)$ is the Heaviside step function, $\mathbf{P}\!\!=\!\!(p_x,p_y,p_z)$ and $\mathbf{R}\!\!=\!\!(x,y,z)$ denote momentum and position, respectively, $V(\mathbf{R})$ is the harmonic trap potential with trapping frequencies $(\omega_x,\omega_y,\omega_z)\!\!=\!\!2\pi\times(11,  15.5, 19)\,$Hz, $g\!=\!4\pi\hbar^2a_s/m$ is the interaction coupling parameter, and $V_{\rm kick}(z)\!=\!V_0\sin^2(k_Lz)$ represents the lattice potential for the QKR. The density is given by $n(\mathbf R, t)\!=\!{\abs{\psi(\mathbf{R},t)}}^2$.\\

The numerical simulation of the 3D QKR with an interacting BEC is done by evolving the 3D GPE using the Hamiltonian given in Eq.\ref{eq_H} with a Fourier-transform-based split-operator time-evolution method. The real-space system sizes in all three dimensions are $\{L_x,L_y,L_z\}\!=\!\{80,80,200\}\,\upmu\textrm{m}$, and grid size in real space is $\{N_x,N_y,N_z\}\!=\!\{2^5,2^5,2^{12}\}$. The time step of evolution is $dt\!=\!T/155$. The phase diagram is obtained by averaging $E$ in the range of $N_\mathrm{p}\!=\!400-500$ for different parameters. The system and grid sizes for the simulation results shown in the insets of Fig.\ref{fig4} are $\{L_x,L_y,L_z\}\!=\!\{80,80,800\}\,\upmu\textrm{m}$ and $\{N_x,N_y,N_z\}\!=\!\{2^5,2^5,2^{15}\}$, respectively.
\\

\subsection{Testing for interactions}\label{breathing}
\begin{figure}[htbp]
    \centering
    \includegraphics[width=1\columnwidth]{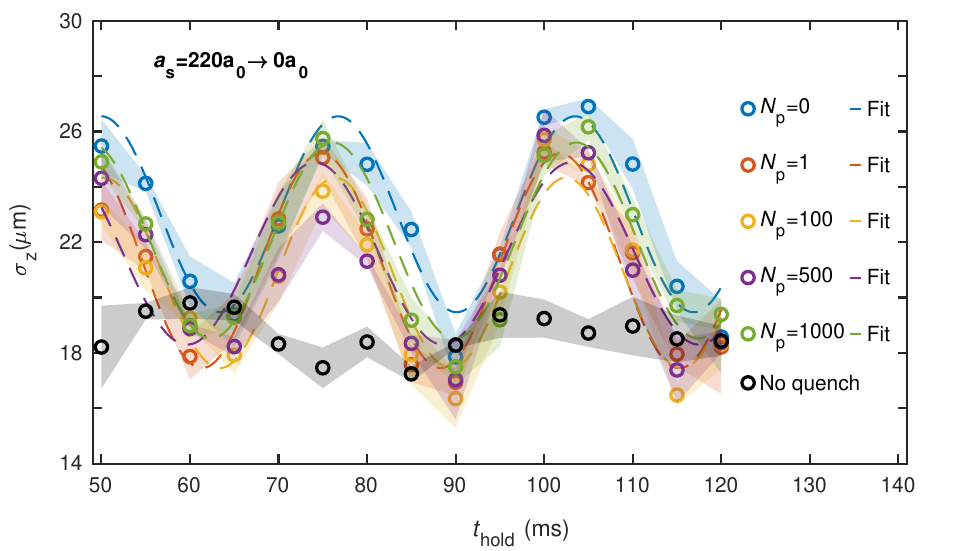}
    \caption{{\bf Interaction quench.} The width of the central peak $\sigma_z$ is plotted as a function of hold time $t_{\text{hold}}$ for initial $a_s\!=\!220\,a_0$ and $\kappa\!=\!0.75$. The colored (grey) data are taken with (without) an interaction quench to $a_s\!=\!0\,a_0$ after $N_\mathrm{p}$ kicks as indicated, ranging from zero to $1000$. The dashed lines are sinusoidal fits to the oscillation data, giving $37.4(4)$ Hz after zero kicks and 36.3(3) Hz after $1000$ kicks. For comparison, the trap frequency along the z-direction is determined to $\omega_z/(2\pi)\!=\! 18.7(3)$ Hz via dipole-mode excitation measurements. The shaded areas reflect the standard deviation obtained from 3 repetitions.}
    \label{fig:breathing}
\end{figure}

Here, we probe the effect of interactions at the final density after the system has undergone a variable amount of kicks. A simple test is to excite the BEC's breathing mode by quenching interactions to zero. Specifically, for the localized phase with $(\kappa, a_s/a_0)\!=\!(0.75, 220)$ we stop the kicking procedure at various values of $N_\mathrm{p}$ and step $a_s$ to $a_s\!=\!0$ within $1$ms. The atoms are then held in the trap for a variable time $t_{\mathrm{hold}}$ before being released from the trap for free expansion with $t_{\textrm{\tiny ToF}}\!=\!40$ ms. The resulting breathing-mode oscillations for the width of the central momentum peak along the $z$-direction are shown in Fig.\ref{fig:breathing}. The measured frequencies match well the value of twice the trap frequency along the vertical direction. In addition, the amplitudes of the oscillations do not change substantially as $N_\mathrm{p}$ is varied, indicating that the interaction energy remains constant during the kicking procedure. This result also confirms that our system has remained localized in real space. 



\subsection{Finite-time scaling}\label{scaling}

We employ a finite-time scaling analysis analogous to finite-size scaling methods~\cite{ardourel_finite-size_2023, PhysRevA.80.043626} to characterize the critical behavior near the localization-delocalization transition. We postulate that the time evolution of $E$ in the critical regime follows a generalized homogeneous function
\begin{equation}
E(t, \kappa) = t^{k_1} \mathcal{F}\left[(\kappa-\kappa_c)t^{k_2}\right],
\end{equation}
where $\mathcal{F}$ is a universal scaling function.  The exponents $k_1$ and $k_2$ are constrained by the requirement that the scaling law recovers the correct asymptotic behavior in both limits. In the localized regime, the energy proxy saturates as $E \sim \ell^2 \sim |\kappa-\kappa_c|^{-2\nu}$, where $\ell$ is the localization length whose critical exponent is $\nu$. In the diffusive regime, it grows as $E \sim D t \sim |\kappa-\kappa_c|^s t$, where $D$ is the diffusion coefficient and $s$ is the critical exponent for conductivity. According to Wegner’s scaling theory~\cite{wegner_electrons_1976}, these exponents obey the relation $s \!=\! (d-2)\nu$, where $d$ is the effective dimensionality of the system. Consistency with these limits requires $k_1 \!=\! 2/d$ and $k_2 \!=\! 1/d\nu$. 
Setting $d\!=\!3$, we get
\begin{equation}
\Lambda(\kappa, t) \equiv E t^{-2/3} = f\left[ \xi(\kappa) t^{-1/3} \right],
\label{eq:scaling_funct}
\end{equation}
where, the scaling parameter $\xi(\kappa) \sim |\kappa-\kappa_c|^{-\nu}$ represents the localization length.

To verify the scaling hypothesis, we collapse the data for varying kick strengths onto a single universal curve. Taking the logarithm of Eq. \ref{eq:scaling_funct} transforms the scaling parameter into a horizontal shift $s_i \!=\! \ln \xi(\kappa_i)$ for each kick strength $\kappa_i$.   Optimal data collapse is achieved by determining the set of horizontal shifts $\{s_i\}$ that minimizes the global variance of the scaled trajectories. By partitioning the $\ln (\Lambda)$ axis into $B$ bins, we define a cost function 
\begin{equation}
\mathcal{C}({s_i}) = \sum_{b=1}^B \sum_{i=1}^M \sum_{j\in S_{ib}} [(X_{ij} + s_i) - \mu_{b}]^2,
\end{equation}
where $X_{ij}$ denotes the unshifted coordinate of the $j$-th point for kick strength $\kappa_i$, $M$ is the number of distinct kick strengths, and $S_{ib}$ represents the subset of points from kick strength $\kappa_i$ falling into bin $b$. The term $\mu_b$ corresponds to the mean position of all shifted data points within the $b$-th bin. We minimize $\mathcal{C}$ using the Nelder-Mead optimization algorithm to retrieve the set $\{s_i\}$ that collapses the trajectories, as shown in Fig. \ref{fig2}d. Finally, as shown in Fig. \ref{fig2}e, we extract the critical kick strength $\kappa_c$ by fitting the resulting scaling parameters $\xi(\kappa)$ to the power law
\begin{equation}
\frac{1}{\xi(\kappa)} = \alpha |\kappa - \kappa_c|^\nu + C,
\end{equation}
where $\alpha$ is a critical amplitude and $C$ is an offset that accounts for experimental effects that smoothen the divergence of the localization length.

\begin{figure}[h!]
\centering
\renewcommand{\figurename}{Fig}
\includegraphics[width=0.85\linewidth]{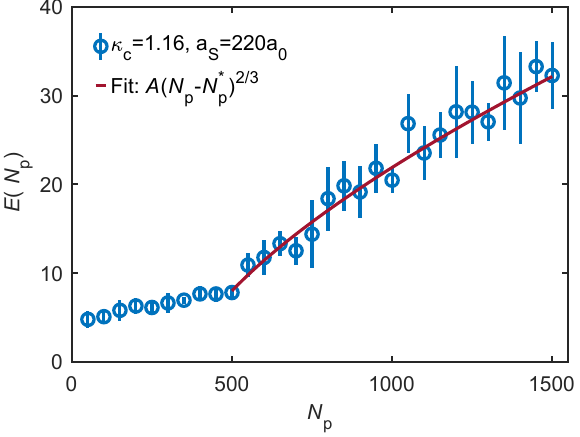}
\caption{{\bf Evolution of the energy proxy at the critical point}: Time evolution of  $E(N_\mathrm{p})$ precisely at the extracted critical kick strength $\kappa_c \!=\! 1.16$. Blue circles represent the experimental data. The late-time data ($N_\mathrm{p} > 500$) are fitted to the shifted power law $E(N_\mathrm{p}) \!=\! A(N_\mathrm{p} - N_\mathrm{p}^*)^{2/3}$ (red solid line). }
\label{fig:tev_critical}
\end{figure}

 We examine the evolution of the energy proxy  precisely at the extracted critical point $\kappa_c \!=\! 1.16$ (see Fig.~\ref{fig:tev_critical}e) to verify the assumption of setting $d\!=\!3$. At criticality, the scaling function becomes a constant, yielding $E(t,\kappa_c) \!=\! t^{2/3}\mathcal{F}(0) \equiv At^{2/3}$. The evolution is therefore scale-invariant and follows a pure power-law growth. To isolate the universal scaling dynamics from the initial, non-universal transient evolution~\cite{gazo2025universal}, we introduce a time offset by setting $t \!=\! N_\mathrm{p} - N_\mathrm{p}^*$. As shown in Fig.~\ref{fig:tev_critical}, the dynamics exhibits a distinct change in behavior around $N_\mathrm{p} \!=\! 500$. Fitting the late-time data ($500 < N_\mathrm{p} \leq 1500$) yields excellent agreement with the predicted power-law exponent of $2/3$ and extracts a transient offset of $N_\mathrm{p}^* \!=\! 358(23)$. The observed critical growth provides an internal consistency check of the choice of effective dimensionality $d\!=\!3$.

A detailed analysis exploring the effective dimensionality, extending the scaling analysis to varying interaction strengths, and determining critical exponents will be presented in a subsequent work~\cite{Karthick}.

\cleardoublepage
\bibliography{ref1}
\cleardoublepage

\end{document}